%% file: corsico.tex
\documentclass[10pt, a4paper, twocolumn, twoside]{article} % 10pt font size (11 and 12 also possible), A4 paper (letterpaper for US letter) and two column layout (remove for one column)

\input{structure.tex} % Specifies the document structure and loads required packages

\title{Blue Large-Amplitude Pulsators (BLAPs): possible origin, evolutionary
  status, and nature of their pulsations} % The article title

\shorttitle{BLAPs} % The short article title for page headings

\shortauthors{C\'orsico, Romero, Althaus, Pelisoli, Kepler} % The short author list for page headings

\author{
  \authorstyle{A.~H.~C\'orsico,$^1$ A.~D.~Romero,$^2$ L.~G.~Althaus,$^1$
    I.~Pelisoli,$^2$ and S.~O.~Kepler$^2$}
	\newline\newline % Space before institutions
	$^1$\institution{Facultad de Ciencias Astron\'omicas y Geof\'isicas, Universidad
          Nacional de La Plata, Paseo del Bosque s/n, (1900) La Plata, Argentina; acorsico,althaus@fcaglp.unlp.edu.ar}\\ % Institution 1
	$^2$\institution{Departamento de Astronomia, Universidade Federal do Rio Grande do Sul, 
          Av. Bento Goncalves 9500, Porto Alegre 91501-970, RS, Brazil; alejandra.romero,ingrid.pelisoli@ufrgs.br,kepler@if.ufrgs.br}\\ % Institution 2
      }

\begin{document}

\maketitle % Print the title

\thispagestyle{firstpage} % Apply the page style for the first page

%----------------------------------------------------------------------------------------
%	ABSTRACT
%----------------------------------------------------------------------------------------

\newabstract{The Blue Large-Amplitude Pulsators (BLAPs)
  constitute a new class of pulsating stars.
  They are hot stars with effective temperatures of $T_{\rm eff}\sim 30\,000$ K
  and surface gravities of $\log g \sim 4.9$, that pulsate with periods in the
  range $\sim 20-40$ min. In \cite{2018MNRAS.477L..30R}, we proposed
  that BLAPs are hot low-mass He-core pre-white dwarf (WD) stars
  that pulsate either in high-order non-radial $g$(gravity) modes or
  low-order radial modes, including the fundamental radial mode.
  The theoretical modes with periods in the observed range are unstable due
  to the $\kappa$ mechanism associated with the $Z$ bump in the opacity at
  $\log T \sim 5.25$. In this work, we extend the study of \cite{2018MNRAS.477L..30R}
  by assessing the rate of period changes of nonradial $g$ modes and radial
  modes and comparing them with the values measured for BLAPs, in an attempt to
  validate the proposed evolutionary scenario, and to discern whether the observed
  modes are high-order $g$ modes or radial modes.}

%----------------------------------------------------------------------------------------
%	ARTICLE BODY
%----------------------------------------------------------------------------------------

\section{Introduction}
\begin{figure*}[t]
  \centerline{\includegraphics[width=2.00\columnwidth]{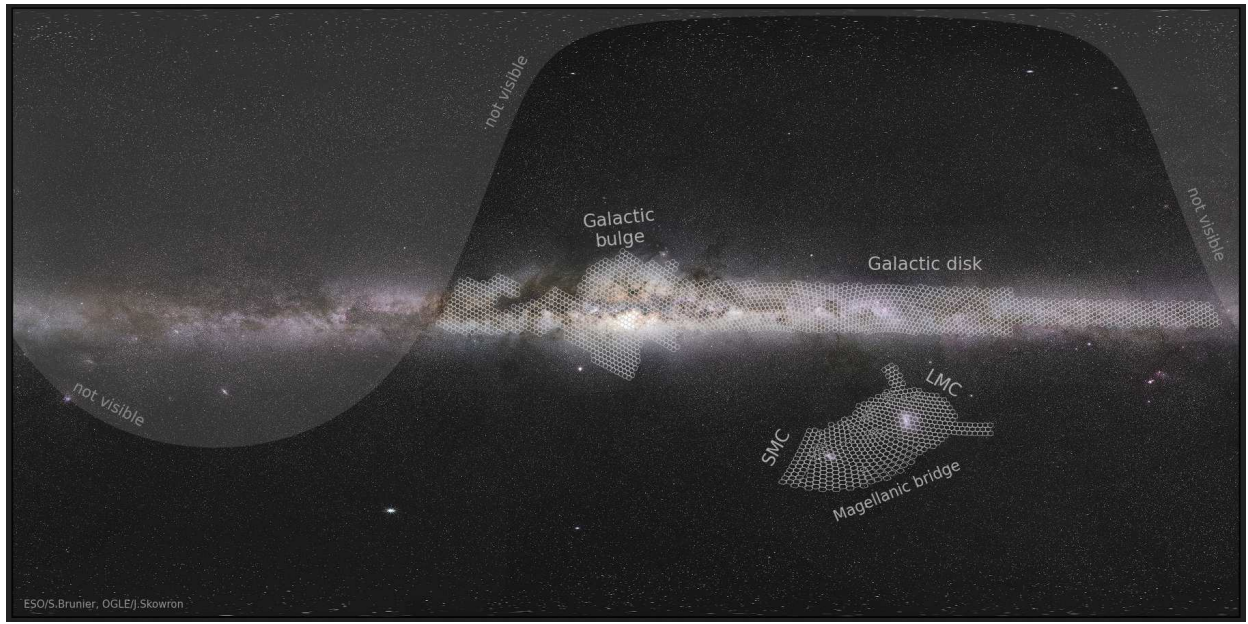}}
  \caption{The sky map with OGLE-IV fields, taken from
{\tt http://www.astrouw.edu.pl/\textasciitilde jskowron/ogle4-sky/}.} 
  \label{fig1}
\end{figure*}

Blue Large-Amplitude Pulsators \citep[BLAPs;][]{2017NatAs...1E.166P} constitute a new
class of pulsating stars recently discovered in the context
of the Optical Gravitational Lensing Experiment \citep[OGLE; ][]{2015AcA....65....1U}.
The first BLAP star (OGLE-BLAP-001) was first tentatively classified as a $\delta$
Scuti-type variable star and named OGLE-GD-DSCT-0058. At present,
14 BLAPs are known. Their average apparent magnitudes are $V= 17.71$ mag and
$I= 17.22$ mag. The fourth phase of OGLE monitors photometrically the Galactic bulge,
the Galactic disc, and the Magellanic Clouds (see Fig. \ref{fig1}). BLAPs have been
discovered only in the Galactic disk and bulge (high-metallicity environments),
but not in the Magellanic Clouds, which constitute a low-metallicity environment
\citep{2018arXiv180204405P}. These variables are not observed in globular
clusters nor in the Galactic halo.  All this suggests high metallicity in these
pulsating stars, something that is confirmed by our theoretical models.

A number of properties observed in BLAPs make these stars striking: 

\begin{itemize}
\item They are very hot stars, with an average effective temperature of
  $T_{\rm eff}\sim 30\,000$ K. Their effective
  temperature and colour change over a complete pulsation cycle, confirming
  that their variability is due to genuine pulsations;
\item The lightcurves of BLAPs have  a saw-tooth shape, reminiscent
  of Cepheids and RR Lyrae-type stars that pulsate in the radial fundamental mode;
\item The amplitudes are large, in the range $0.2-0.4$ mag, much larger than 
  the amplitudes of pulsating sdB stars, $\beta$ Cephei stars,
  Slowly Pulsating B (SPB) stars, etc, which exhibit amplitudes of mmag (milli magnitudes);
\item BLAPs exhibit single short pulsation periods ($\Pi$) in the
  range $\sim 1200-2400$ sec. These periods are much shorter than those
  of  $\beta$ Cephei stars, SPBs, etc. The fact that only a single period has been detected up to date
  in BLAPs does not mean that these stars are not multi-periodic stars. Indeed, they could
  be pulsating with other periods that have not been detected due to an insufficient observation
  time. 
\item The periods of BLAPs show a secular drift with typical values
  of the relative rate of period change of $\dot{\Pi}/\Pi=d(\log \Pi)/dt= 10^{-7}$ yr$^{-1}$,
  both positive (increasing periods) and negative (decreasing periods). The magnitudes of the
  rate of period change suggest that BLAPs are stars that are slowly evolving
  on nuclear timescales;  
\item BLAPs exhibit envelopes made of a mixture of H and He. 
\end{itemize}

\section{A possible evolutionary origin of BLAPs}

\begin{figure}[t]
  \centerline{\includegraphics[width=1.0\columnwidth]{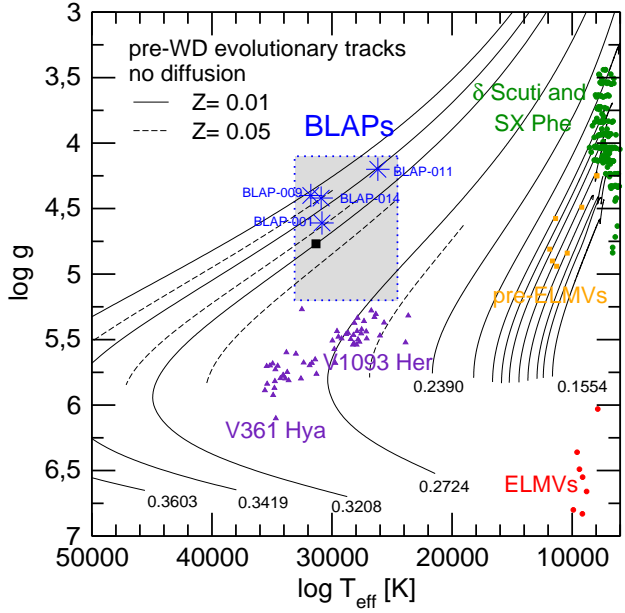}}
  \caption{$T_{\rm eff} - \log g$ diagram showing the
    location of the BLAP stars (shaded rectangle area),
    along with other families of already known
    pulsating stars: ELMVs (red dots), pre-ELMVs (orange dots),
    pulsating sdBs (V361 Hya and V1093 Her; violet triangles) and
    $\delta$ Sct/SX Phe stars (green dots).
    Solid black lines correspond to low-mass He-core pre-WD evolutionary
    tracks computed neglecting element diffusion and $Z= 0.01$.
    Numbers correspond to the stellar mass of some sequences.
    Also included are portions of evolutionary tracks corresponding to
    $Z= 0.05$ for some stellar masses. Blue star symbols indicate the location of
    the four BLAP stars with measured atmospheric parameters. The black
    square on the evolutionary track of $M_{\star}= 0.3208 M_{\odot}$ indicates the
    location of a template model.} 
  \label{fig2}
\end{figure}

Any evolutionary/pulsational model proposed to explain the existence of BLAPs and
the nature of their pulsations must satisfy the observational constraints listed in the previous
section. 
In the case of 4 BLAPs, it  was possible to derive their $T_{\rm eff}$ and
$\log g$ from spectroscopy \citep{2017NatAs...1E.166P}. These stars are
located in a region in the $T_{\rm eff}-\log g$ diagram that is not occupied by any
previously known kind of pulsating stars, as depicted in Fig. \ref{fig2}. 
Indeed, BLAPs have similar gravities as pre-ELMVs, but are hotter than them;
they are much hotter and more compact than $\delta$ Scuti/SX Phe stars; they have similar
$T_{\rm eff}$s but are less compact than pulsating sdBs (V361 Hya and V1093 Her types);
finally, they are much hotter and less compact than ELMVs. In view of the
above-mentioned properties, the following questions arise: what is the internal
structure and the evolutionary status of BLAPs? what is their evolutionary origin?
Several possibilities to explain the formation, evolution and internal structure of these
stars have been proposed \citep{2017NatAs...1E.166P}. On one hand, the evolution of
a single isolated low-mass stars seems impossible to explain BLAPs, because the
evolutionary timescales involved in such a scenario are longer than the Hubble time. 
On the other hand, binary-star evolution through stable mass transfer and/or common
envelope ejection appear as more plausible evolutionary channels for these intriguing
pulsating stars. \cite{2017NatAs...1E.166P} proposed two main possibilities:

\begin{itemize}
\item He-core shell H burning  low-mass stars ($\sim 0.30 M_{\odot}$)
\item core He-burning stars ($\sim 1.0 M_{\odot}$)  
\end{itemize}  

In \cite{2018MNRAS.477L..30R}, we proposed that BLAPs are
hot He-core shell H burning low-mass pre-WD stars with masses $\sim 0.30 M_{\odot}$.
In Fig. \ref{fig2} we display the the evolutionary tracks for
low-mass He-core pre-WD models of \cite{2013A&A...557A..19A} which neglect
element diffusion, corresponding to solar metallicity ($Z= 0.01$) and
super-solar metallicity ($Z= 0.05$). Clearly, the location of the BLAPs is well
accounted for by these evolutionary tracks, demonstrating that the scenario
proposed by \cite{2018MNRAS.477L..30R} is plausible. 

\section{Nature of the pulsations of BLAPs}

\begin{figure}[t]
  \centerline{\includegraphics[width=1.0\columnwidth]{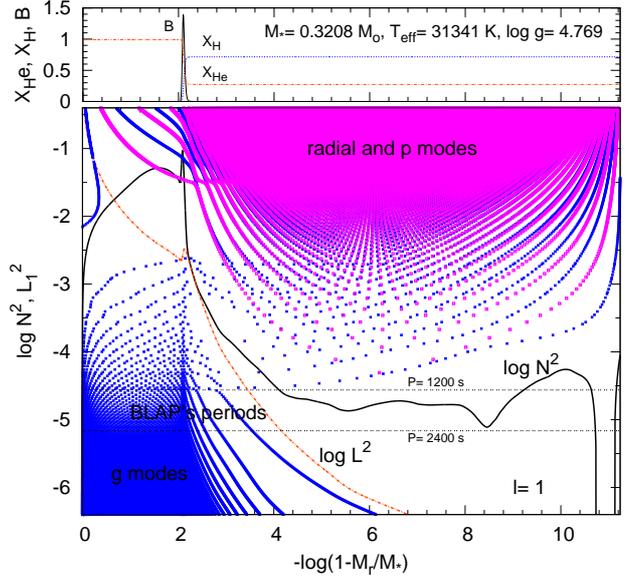}}
  \caption{Internal chemical profiles for He and H and the Ledoux
    term B (upper panel) and the propagation diagram (lower panel),
    in terms of the outer mass fraction coordinate,
corresponding to the pre-ELM WD template model of $M_{\star}= 0.308M_{\odot}$
and $T_{\rm eff} \sim 31\,300$ K marked in Fig. \ref{fig2} with a black square.
In the lower panel, tiny star symbols (in blue) correspond to the spatial
location of the nodes of the radial eigenfunction of nonradial dipole ($\ell= 1$)
$g$ and $p$ modes. Tiny squares (in magenta) mark the location of the nodes for
radial ($\ell= 0$) modes. The (squared) frequency interval
corresponding to the modes observed in BLAPs (with periods
in the range $\sim 1200-2400$ s) is emphasized with two horizontal
black dotted lines.} 
\label{fig3}
\end{figure}

In Fig. \ref{fig3}, we show the internal chemical profiles and the Ledoux term
$B$ (that is crucial in the computation of the Brunt-V\"ais\"al\"a frequency) in terms
of the outer mass fraction coordinate (upper panel), and a \emph{propagation diagram}
---the run of the logarithm of the squared critical frequencies, that is, $N^2$
(the Brunt-V\"ais\"al\"a frequency), and $L_{\ell= 1}^2$ (the Lamb frequency) ---
with the nodes of the radial eigenfunction for nonradial ($\ell= 1$) $g$ and $p$ modes
and also radial ($\ell= 0$) modes (lower panel), corresponding to a template model with 
$M_{\star}= 0.3208 M_{\odot}$, $T_{\rm eff}= $ and $Z= 0.01$ whose location
in the $T_{\rm eff}-\log g$ diagram is marked in Fig. \ref{fig2} with a black square.
The pulsation computations were carried out with the adiabatic version of the
{\tt LP-PUL} pulsation
code \citep{2006A&A...454..863C}. For this specific template model,  high-order $g$ modes with
radial orders $k= 25-50$ and  radial modes with the lowest radial order ($k= 0$)
have pulsation periods in the observed interval ($\sim 1200-2400$ s). Therefore, the
periodicities exhibited by BLAPs can be associated either to nonradial $g$ modes,
or to radial modes \citep{2018MNRAS.477L..30R}.

Linear nonadiabatic pulsation computations performed with the nonadiabatic version of the
{\tt LP-PUL} pulsation code \citep{2006A&A...458..259C} do not predict pulsations 
for models with $Z \leq 0.03$. However, for higher metallicities, unstable radial and
nonradial modes with periods compatible with those observed in BLAPs
are found. In particular, for a template model with $Z= 0.05$ ($M_{\star}= 0.3419 M_{\odot}$
and $T_{\rm eff}= 31\,100$ K) we found that high-order $g$
modes with radial orders $k= 25-39$ $(\ell= 1)$ and $k= 46-67$ $(\ell= 2)$, and
radial modes with low radial order $k= 0$ (fundamental mode) are destabilized by the
$\kappa$ mechanism due to the ``$Z$ bump'' in the Rosseland
opacity at $\log T \sim 5.25$ due to
Fe (and the iron group) like in $\beta$ Cephei stars, SPBs, and variable sdB stars.
It is worth mentioning that the He$^{++}$ bump in the opacity
at $\log T \sim 4.45$, which is located at more external regions of the star,
is unable to destabilize pulsations, unlike  what happens in the case of pre-ELMVs 
\citep{2016A&A...588A..74C}.

\begin{figure}[t]
  \centerline{\includegraphics[width=1.0\columnwidth]{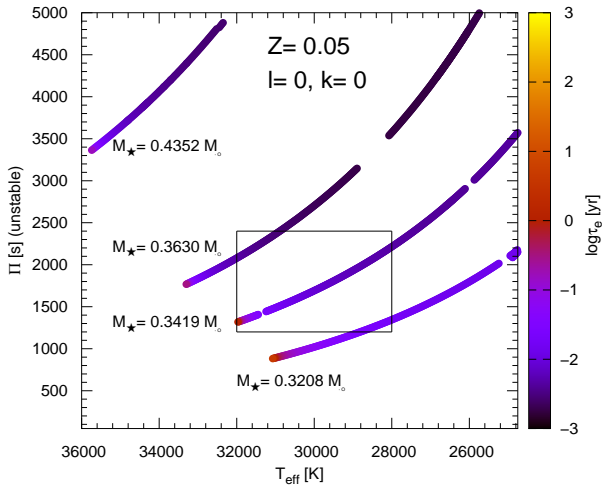}}
  \caption{Periods of unstable fundamental ($k= 0$) radial mode ($\ell= 0$)
    in terms of $T_{\rm eff}$ for pre-WD models of the indicated masses and $Z= 0.05$.
    The palette of colors at the right scale indicates the
    value of the logarithm of the $e$-folding time (in yrs). The $e$-folding
    times range from $\sim 10^{-3}$ to $\sim 10^3$ yr, much shorter than the typical
    evolutionary timescales at that stage of evolution. The
    rectangle corresponds to the interval of effective temperature
    measured for BLAPs and the range of detected periods.}
  \label{fig4}
\end{figure}

\begin{figure}[t]
  \centerline{\includegraphics[width=1.0\columnwidth]{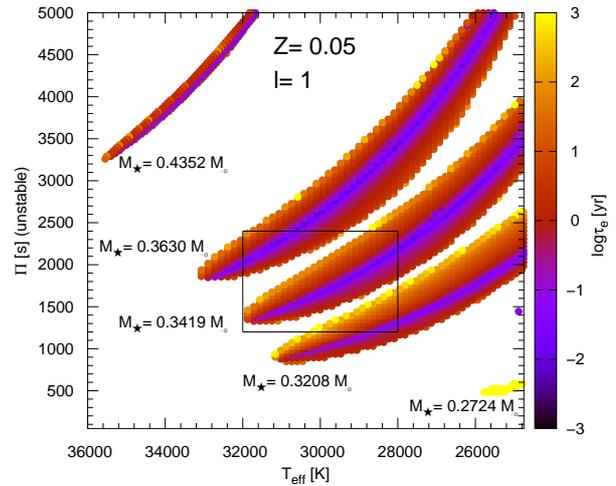}}
  \caption{Same as in Fig. \ref{fig4}, but for the case of nonradial $g$ modes
  with $\ell= 1$ and a range of radial orders $k$.}
  \label{fig5}
\end{figure}

\begin{figure}[t]
  \centerline{\includegraphics[width=1.0\columnwidth]{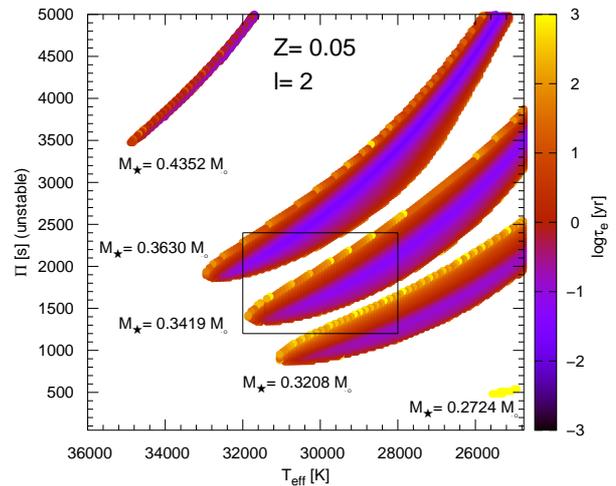}}
  \caption{Same as in Fig. \ref{fig5}, but for the case of $\ell= 2$.}
  \label{fig6}
\end{figure}

By extending the stability calculations to evolutionary sequences with different
stellar masses and covering the range of effective temperatures of interest,
it is possible to define the complete domains of instability of BLAPs
in the $T_{\rm eff} - P$ diagrams. We show the results in Figs. \ref{fig4},  \ref{fig5}, and
\ref{fig6}, corresponding to unstable modes with $\ell= 0$, $\ell= 1$, and $\ell= 2$,
respectively. In the case of radial modes (Fig. \ref{fig4}), only the fundamental mode is
unstable. Note that the observed periodicities $(1200 \leq P \leq 2400)$ s
are well accounted for by the theoretical computations
if we consider a range of stellar masses, $0.33 \leq M_{\star}/M_{\odot} \leq 0.36$.
It is apparent that the radial fundamental modes are the most unstable ones
among the studied cases ($\ell= 0, 1, 2$). Indeed, they are destabilized during very short
times ($e$-folding times) as compared with the evolutionary timescales at that
stage of evolution. The results for $\ell= 1$
and $\ell= 2$ are virtually the same, i.e., the domains of instability for nonradial modes
do not depend on the value of $\ell$.

At this point, and considering the results of our analysis, it seems that
the variability of the BLAPs is better explained by the possible excitation of the
fundamental radial mode in these stars, rather than by the high-order $g$ modes,
for a number of reasons: {\it (i)} the fundamental radial mode has the correct period,
{\it (ii)} it is pulsationally unstable in the range of effective temperatures
of interest, {\it (iii)} it is more unstable than the nonradial $g$ modes,
and {\it (iv)} the fact that BLAPs exhibit a single mode with large amplitude
in the lightcurves, reminiscent to a radial mode. In the next Section,
we analyze in detail the sign and magnitude of the theoretical rates of period change
of our models for radial and nonradial modes, and compare them with the values
measured in BLAPs.

%------------------------------------------------

\section{Rate of period changes of BLAPs}

\begin{figure}[t]
  \centerline{\includegraphics[width=1.0\columnwidth]{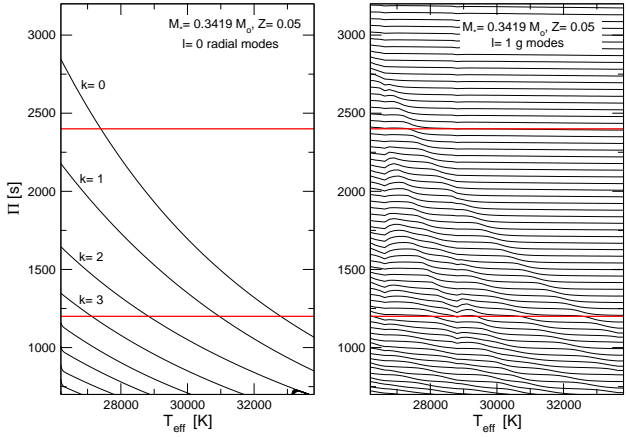}}
  \caption{The pulsation periods of radial modes
    (left panel) and nonradial  $g$ modes with $\ell= 1$ (right panel) in terms of
    the effective temperature, for a sequence of models with $M_{\star}= 0.3419 M_{\odot}$
    and $Z= 0.05$. The red horizontal lines correspond to the limits of the period
    interval observed in BLAPs.}
  \label{fig7}
\end{figure}

In general, the rates of period change in pulsating stars are associated to 
their evolutionary timescale. In the case of BLAPs, \cite{2017NatAs...1E.166P} 
(their Table 1) have measured the relative rates of period change (positive and negative)
of 11 stars with an average absolute value of
$|\dot{P}/P| \sim 10^{-7}$ yr$^{-1}$, and extreme values of $(+7.65\pm0.67) \times 10^{-7}$
yr$^{-1}$ and
$(-2.85 \pm 0.31) \times 10^{-7}$ yr$^{-1}$. These are relatively large values of the
rates of period change, which suggest that BLAPs are slowly evolving on nuclear
timescales. We have computed the rates of period change for our He-core shell H burning 
pre-WD models for the cases of radial and and nonradial modes. In
Fig. \ref{fig7} we plot the evolution with $T_{\rm eff}$ of the pulsation periods for
radial modes ($\ell= 0$, left panel) and nonradial $\ell= 1$  $g$ modes (right panel),
corresponding to an evolutionary sequence with $M_{\star}= 0.3419 M_{\odot}$ and $Z= 0.05$. 
In general, the slope of the periods of radial modes is much larger than
in the case of $g$ modes, indicating a larger rate of change of periods in the case
of radial modes. On the other hand, all the periods of radial modes decrease with
increasing $T_{\rm eff}$, showing that  the rates of change of these periods must
be all negative. In contrast, in the case of the $g$ modes, there are parts
of the evolution where the periods decrease and other parts where the periods grow,
indicating that negative and positive values of the rates of period changes are
expected for $g$ modes.

\begin{figure}[t]
  \centerline{\includegraphics[width=1.0\columnwidth]{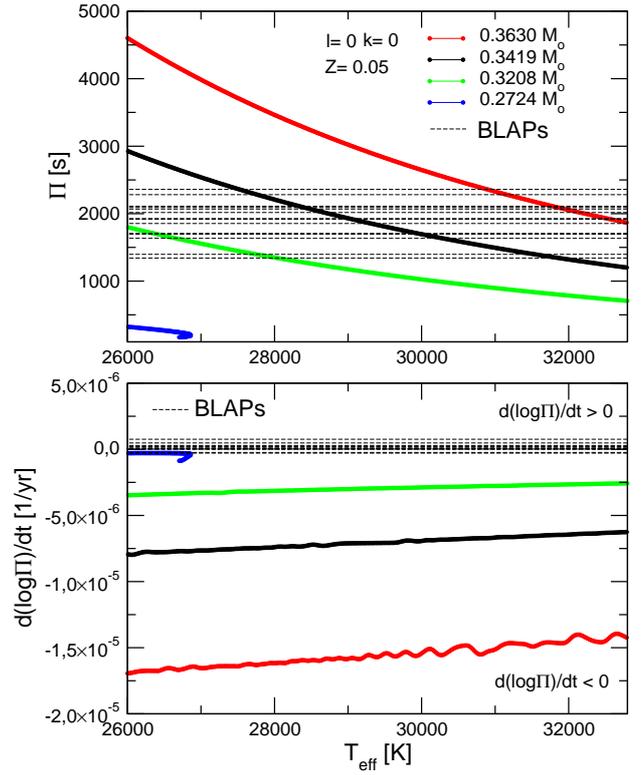}}
  \caption{Upper panel: the periods of radial fundamental mode ($\ell= 0, k= 0$) versus
    the effective temperature, for model sequences with different stellar masses and
    $Z= 0.05$. Horizontal dashed lines are the periods observed in BLAPs.
    Lower panel: same as upper panel but for the case of the relative rates of period
    changes. The horizontal dashed lines are the values of $\dot{\Pi}/\Pi$ measured in
    BLAPs.}
  \label{fig8}
\end{figure}

In the upper and lower panels of Fig. \ref{fig8} we depict the period ($\Pi$)
and the relative rate of period change ($\dot{\Pi}/\Pi$), respectively,
for the radial fundamental mode ---the only radial mode that is predicted to be
pulsationally unstable according to our
nonadiabatic computations--- in terms of the effective temperature, for model sequences
with different stellar masses and $Z= 0.05$. Horizontal dashed lines correspond to
the periods (upper panel) and the relative rates of period changes (lower panel)
measured for BLAPs. Note that the radial fundamental mode ($\ell= 0, k= 0$)
for different values of $M_{\star}$ has a period in excellent agreement
with the periods observed in BLAPs in the
range of effective temperature in which these stars are found, except in the
case of $M_{\star}= 0.2724 M_{\odot}$. On the other hand, from the lower panel of the
figure is clear that the absolute value of the rate of period change for the
fundamental mode ($|\dot{\Pi}/\Pi| \sim 10^{-5}-10^{-6}$ yr$^{-1}$) is largely in excess when
compared with the values measured for BLAPs ($\sim 10^{-7}$ yr$^{-1}$), except
in the case of $M_{\star}= 0.2724 M_{\odot}$. In addition, all
the values are negative, as anticipated in the left panel of Fig. \ref{fig7}.
We conclude that, according to the values of the rate of period change measured for
BLAPs, the pulsations of all these stars can not be attributed to the 
radial fundamental mode (nor to radial modes in general). 

\begin{figure}[t]
  \centerline{\includegraphics[width=1.0\columnwidth]{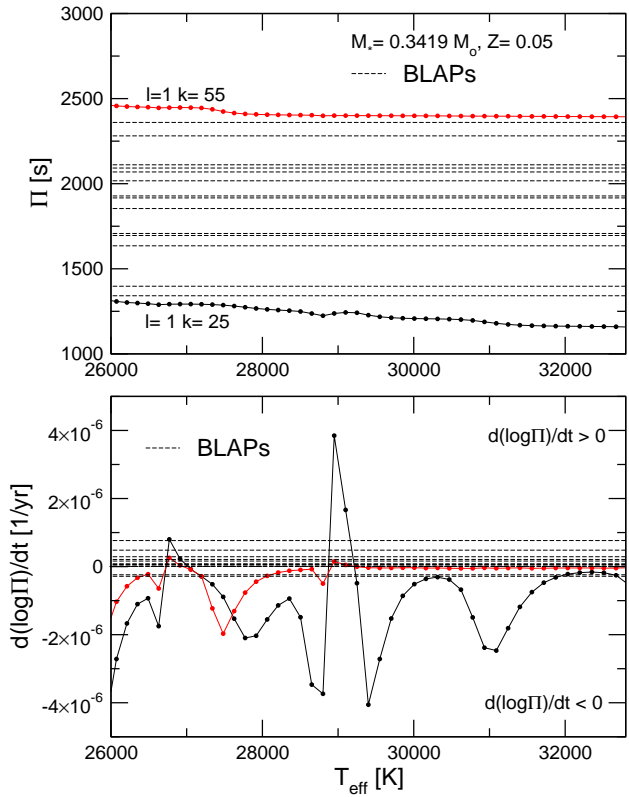}}
  \caption{Upper panel: the periods of the nonradial mode ($\ell= 1$) with $k= 25$
    (black dotted line) and
    $k= 55$ (red dotted line) in terms of $T_{\rm eff}$, for a model
    sequence with $M_{\star}= 0.3419 M_{\odot}$
    and $Z= 0.05$. Horizontal dashed lines are the periods observed in BLAPs.
    Lower panel: same as upper panel but for the case of the relative rates of period
    changes. The horizontal dashed lines are the values of $\dot{\Pi}/\Pi$ measured in
    BLAPs. Note that $|\dot{\Pi}/\Pi| < 4 \times 10^{-6}$ yr$^{-1}$ for
    models.}
  \label{fig9}
\end{figure}

What about nonradial $g$ modes? In the upper panel of Fig. \ref{fig9} we show the periods 
of the $\ell= 1$ $g$ modes with $k= 25$ and $k= 55$ for the sequence with
$M_{\star}= 0.3419 M_{\odot}$ and $Z= 0.05$. These periods are close to the limits
of the interval of periods measured in the BLAPs (dashed horizontal lines). In the upper
panel of the figure we depict the corresponding values of $\dot{\Pi}/\Pi$. At variance
with what happens for the  fundamental radial mode (lower panel of Fig. \ref{fig8}),
in this case the rates of period change are positive and negative, and for certain
ranges of $T_{\rm eff}$, they adopt values compatible with those measured in BLAPs
(dashed horizontal lines). Thus, we can conclude that the rates of period
changes exhibited by BLAPs could be satisfactorily explained by high-order nonradial $g$
modes.

\section{Conclusions}

In this work, we have investigated a possible evolutionary origin for the BLAP
stars as being the hot counterpart of the already known pre-ELMV stars
---pulsating low-mass He-core shell H burning pre-WD stars, \citep{2016A&A...588A..74C}.
According to this evolutionary scenario, BLAPs could be $\sim 0.3 M_{\odot}$ pre-WDs
at $T_{\rm eff} \sim 30\,000$ K with He/H envelopes resulting from binary-star
evolution \citep[see][for details]{2018MNRAS.477L..30R}. If this scenario were correct,
the companion star should be observed.
However, no companion star is observed in any BLAP\footnote{We note that
  in sdB stars, about half of the supposed companions are also not observed.}. This 
could be simply due to the faintness of the companion ---no eclipses are seen, no lines of
companions are detected.
Similar investigation should be done for BLAPs. Another alternative is that
BLAPs are the result of
mergers of ELM+ELM binary systems (each component with $\sim 0.15 M_{\odot}$). Finally,
another possibility is that BLAPs are actually core He-burning stars with masses
$\sim 1.0 M_{\odot}$ \citep{2017NatAs...1E.166P}, although in that case, no feasible
evolutionary channel is known. Gaia distance for OGLE-BLAP-009, of about $2.63$ kpc,  
seems to favour the low-mass ($\sim 0.30 M_{\odot}$) solution over the $1.0 M_{\odot}$ solution
(Pawel Pietrukowicz, private communication). We conclude that the evolutionary origin
of BLAPs is still an open problem. 

Regarding the kind of pulsation modes responsible for the variability of BLAPs,
we have obtained two possible interpretations. On one hand, the pulsations
of these stars could be due to the radial fundamental mode ($\ell= 0, k= 0$).
These modes have the right value of the periods at the correct effective temperatures
and gravities (stellar masses), and, in addition, they are strongly destabilized
by the $\kappa$ mechanism due to the $Z$ bump, provided that stellar
models with enhanced metallicity $Z \sim  0.05$ are considered. It is important
to mention that the fundamental radial mode is the only radial mode that is destabilized in our
computations. On the other hand, high-order nonradial $g$ modes also have pulsation
periods in agreement with those observed in BLAPs, at the right effective temperatures 
and gravities. A lot of these modes also are unstable by the $\kappa$ mechanism
due to the $Z$ bump, although they are not so strongly destabilized as the
fundamental radial mode is. In view of this, and taking into account the fact
that in BLAPs just a single period has been detected to date with a large amplitude
in the lightcurves,
it seems that the natural interpretation of the variability in BLAPs should be
the radial fundamental mode. However, when we examine the rates of period
changes of our models, we found that nonradial $g$ modes with high radial order $k$
are characterized by $\dot{\Pi}/\Pi$ values in much better agreement with the
values measured in BLAPs than the fundamental radial mode. In summary,
the exact nature of the pulsation modes responsible for the variability of BLAPs
remains a matter of debate.

Finally, there is the issue of the high metallicity necessary to excite
pulsation modes through the $\kappa$ mechanism. In our models,
the metallicity is \emph{globally} enhanced in order to have an enhanced $Z$ bump in the
Rosseland opacity.
However, it would be possible to found instability with a
\emph{local} enhancement of the opacity at the region of the star in which the $Z$ bump is
located. This could be achieved if radiative levitation is operative, in such a
way that Fe locally accumulates at the driving zone. This problem
needs to be investigated consistently and will be addressed in a future work.

\section*{Acknowledgements} A.H.C. warmly thanks  the Local Organising Committee
of the  21th European White Dwarf Workshop for support that allowed him to
attend this conference. The authors thank Pawel Pietrukowicz for sharing
valuable information about BLAPs.

%----------------------------------------------------------------------------------------
%	BIBLIOGRAPHY
%----------------------------------------------------------------------------------------

% There are two ways to include references. The first uses bibtex and
% is recommended. For this case uncomment the following line. 
\bibliography{papers}

\end{document}

%% file: structure.tex
\usepackage[english]{babel} % English language hyphenation

\usepackage{microtype} % Better typography

\usepackage{amsmath,amsfonts,amsthm} % Math packages for equations

\usepackage[svgnames]{xcolor} % Enabling colors by their 'svgnames'

\usepackage[small, labelfont=bf, up]{caption} % Custom captions under/above tables and figures

\usepackage{booktabs} % Horizontal rules in tables

\usepackage{lastpage} % Used to determine the number of pages in the document (for "Page X of Total")

\usepackage{graphicx} % Required for adding images

\usepackage{enumitem} % Required for customising lists
\setlist{noitemsep} % Remove spacing between bullet/numbered list elements

\usepackage{sectsty} % Enables custom section titles
\allsectionsfont{\usefont{OT1}{phv}{b}{n}} % Change the font of all section commands (Helvetica)

\usepackage{geometry} % Required for adjusting page dimensions

\usepackage[T1]{fontenc} % Output font encoding for international characters
\usepackage[utf8]{inputenc} % Required for inputting international characters

\usepackage{XCharter} % Use the XCharter font

\usepackage{hyperref}

\usepackage{journals}

\usepackage{fancyhdr} % Needed to define custom headers/footers
\newcommand{\shorttitle}[1]{\fancyhead[CE]{\textsl{#1}}}
\newcommand{\shortauthors}[1]{\fancyhead[CO]{\textsl{#1}}}

\fancypagestyle{firstpage}{ % Page style for the first page with the title
	\fancyhf{}
        \lhead{\vspace*{0.0em} \textit{\footnotesize 21$^{\rm st}$ European Workshop on White
            Dwarfs \\ Castanheira, Campos, Montgomery, eds. \\[-0.2em]
          July 23--27, 2018, Austin, Texas, USA}}
}

\date{}

\newcommand{\authorstyle}[1]{{\large\usefont{OT1}{phv}{b}{n}\color{DarkRed}#1}} % Authors style (Helvetica)

\newcommand{\institution}[1]{{\footnotesize\usefont{OT1}{phv}{m}{sl}\color{Black}#1}} % Institutions style (Helvetica)
\usepackage{titling} % Allows custom title configuration

\newcommand{\HorRule}{\color{DarkGoldenrod}\rule{\linewidth}{1pt}} % Defines the gold horizontal rule around the title

\pretitle{
	\vspace{-30pt} % Move the entire title section up
	\HorRule\vspace{10pt} % Horizontal rule before the title
	\fontsize{16}{12}\usefont{OT1}{phv}{b}{n}\selectfont % Helvetica
	\color{DarkRed} % Text colour for the title and author(s)
}

\posttitle{\par\vskip 10pt} % Whitespace under the title

\preauthor{} % Anything that will appear before \author is printed

\postauthor{ % Anything that will appear after \author is printed
	\vspace{0pt} % Space before the rule
	{\par\HorRule} % Horizontal rule after the title
	\vspace{-10pt} % Space after the title section
}

\newcommand{\newabstract}[1]{
    {\section*{Abstract}
    \bfseries #1}
  }

\usepackage[authoryear]{natbib}